\documentclass[aps]{revtex4}
\usepackage{amssymb}
\usepackage[dvips]{graphicx}
\usepackage[english]{babel}
\usepackage{indentfirst}
\usepackage{amsxtra}
\usepackage{amsmath}
\usepackage{supertabular}
\usepackage{multirow}
\usepackage [mathcal]{eucal}
\newcommand {\oxy}{$^{16}$O }
\newcommand {\caI}{$^{40}$Ca }
\newcommand {\caII}{$^{48}$Ca }

\def\beq{\begin{equation}}
\def\eeq{\end{equation}}
\def\beqn{ \begin{eqnarray} }
\def\eeqn{ \end{eqnarray} }
\def\s1s2{{ \boldsymbol{\sigma}(1) \cdot \boldsymbol{\sigma}(2) }}
\def\t1t2{{ \boldsymbol{\tau}(1) \cdot \boldsymbol{\tau}(2)  }}

\newcommand{\nr}{{\bf r}}
\newcommand{\bq}{{\bf q}}

\begin{document}
\noindent
\title{Effective tensor forces and neutron rich nuclei}

\author{G. Co' and V. De Donno}
\affiliation{Dipartimento di Fisica, Universit\`a del Salento and,
 INFN Sezione di Lecce, Via Arnesano, I-73100 Lecce, ITALY}
\author{M. Anguiano and A. M. Lallena}
\affiliation{Departamento de F\'\i sica At\'omica, Molecular y
  Nuclear, Universidad de Granada, E-18071 Granada, SPAIN}

\date{\today}

\begin{abstract}
We study the effects of the tensor term of the effective
nucleon-nucleon interaction on nuclear excited states.  Our
investigation has been conducted by using a self-consistent Random
Phase Approximation approach. We investigate various nuclei in
different regions of the isotopes chart. Results for a set of calcium
isotopes are shown.
\end{abstract}

\maketitle

The tensor parts of the nucleon-nucleon interaction are usually
neglected in Hartree-Fock (HF) and Random Phase Approximation (RPA)
calculations.  In these theories, the effects of the tensor
force are taken into account by modifying the values of the parameters
of the other terms of the effective force. We have investigated the
validity of this approach, and we present here some results of our
study.

In our work we have considered only the tensor-isospin term of the
nuclear interaction, which we have described as
\beq
v^{t\tau}_{b}(r) 
=  v^{t\tau}_{AV8'}(r)  (1 -  e^{-b r^2})
\label{eq:tens}
\,\,\,,
\eeq 
where we have indicated with $v^{t\tau}_{AV8'}(r)$ the tensor-isospin
term of the Argonne V8' microscopic potential \cite{pud97}. In the
above expression, this term is multiplied by a function which
considers the effects of the short-range correlations.  The only free
parameter is $b$, whose value rules the strength of the tensor part of
the force.  Small values of $b$ increase the active range of the
correlation reducing the strength of the force.

\begin{figure}[h]
\includegraphics[width=7.0 cm] {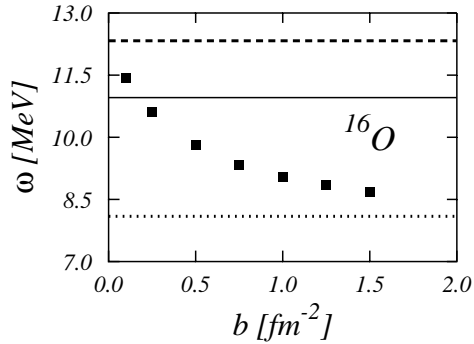} 
\hspace{0.5 cm}   
\begin{minipage}[b]{6.0 cm}  
\caption{
Energy of the first 0$^-$ excited state in \oxy calculated by using a
phenomenological RPA approach. The dashed line indicates the values
obtained without tensor force. 
The dotted line shows the value obtained when the 
full Argonne V8' term is active. The squares indicate the values
obtained for various values of the parameter $b$ of
Eq. \ref{eq:tens}. The full line shows the experimental value.\\~\\
\label{fig:LM}
}
\end{minipage}
\end{figure}

As example of the effects produced by the tensor force on the nuclear
excited states, we show in Fig. \ref{fig:LM} the evolution of the RPA
eigenenergy of the first 0$^-$ excited state in \oxy, obtained with a
phenomenological RPA calculation. Similar sensitivity of the 0$^-$
excitation energies to the presence of the tensor term of the force
has been found in all the nuclei we have investigated, $^{12}$C,
$^{16}$O, $^{40}$Ca, $^{48}$Ca and $^{208}$Pb. We present here only
the \oxy case for sake of simplicity.  The results of
Fig. \ref{fig:LM} have been obtained by using single particle energies
and wave functions calculated by diagonalizing a Woods-Saxon potential
in a harmonic oscillator basis. The residual interaction, used in the
RPA calculation, is a zero-range Landau-Migdal force whose 
parameters values are those given in Ref. \cite{don09}. 
The value obtained with
this interaction is shown in the figure by the dashed line. When the
full Argonne V8' tensor term is added, we obtain the value represented
by the dotted line. The full line shows the experimental value of
10.96 MeV \cite{led78}.  The squares indicate the values of the 0$^-$
eigenenergies as a function of $b$.  These values become smaller with
the increase of the tensor force strength.  The evolution of the 0$^-$
eigenenergy, from that obtained with the pure Landau-Migdal force up
to that obtained when the full tensor term is active, is very smooth.

\vspace*{0.05cm}
The phenomenological RPA calculations described above have been used to
identify an observable very sensitive to the tensor force. We have
considered this quantity in order to select the values of $b$.

The tensor force produces effects on the single particle basis and
also on the excited states of the nucleus. For this reason we have
conducted our study by doing self-consistent RPA calculations. This
means that we used in the RPA calculations the single particle wave
functions and energies obtained by solving the HF equations.  The
effective interaction in both type of calculations is the same.

The effect of the tensor force on the single particle energies has
been well clarified by Otsuka and collaborators \cite{ots05}.  For
example, the energies of two proton spin-obit partner single particle
levels are modified by the tensor-isospin interaction acting with a
neutron single particle level. If the total angular momentum of the
occupied neutron single particle level is $j=l+1/2$ the energy of the
$j=l-1/2$ proton level is lowered and that of the $j=l+1/2$ proton
level is increased.  This effect is reversed when the occupied neutron
level has $j=l-1/2$. For this reason, the nuclei where all the neutron
spin-orbit partners single particle levels are fully occupied do not
show any tensor force effect.

We have conducted our study by comparing results obtained with and
without tensor force.  We have considered two different
parametrization of the Gogny interaction. They are the traditional
D1S \cite{ber91} force and the more modern D1M parametrization
\cite{gor09} which describes better the behaviour of the the
microscopic equation of state of neutron matter at high densities.

The D1S and D1M interactions do not have tensor terms. We add to these
interactions a tensor-isospin term of the form given by
Eq. \ref{eq:tens}.  The procedure we have adopted to construct new
parametrization of the force is iterative. We first made a HF
calculation without tensor force to produce single particle energies and
wave functions. Then we made RPA calculations with the tensor force,
selecting the value of $b$ which reproduces the experimental energy of
the 0$^-$ in $^{16}$O.  With this new interaction we have recalculated
the HF single particle energies and wave functions, and we changed the
spin-orbit term of the interaction to reproduce the splitting between
the neutron 1p$_{3/2}$ and 1p$_{1/2}$ levels.  With this single
particle basis we performed a new set of RPA calculations and the
procedure has been repeated up to convergence. In this way, starting
from the D1S and D1M forces, we have constructed two new
parametrizations which we call respectively D1SV8 and D1MV8.  The
differences with the original parametrizations are only in the
spin-orbit term and, obviously, in the presence of the tensor term. We
did not make a complete refit of the forces.

In our studies, we have also used the GT2 interaction \cite{ots06}
which contains a tensor term, and we made calculations by switching
off this term. Since the results obtained in this way are very similar
to those obtained with the other two interactions, for sake of
simplicity, we do not present them here.

\begin{figure}[h]
\includegraphics[width=7.0 cm] {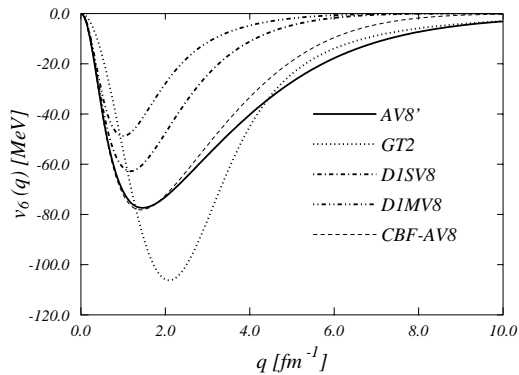} 
\hspace{0.5 cm}   
\begin{minipage}[b]{6.0 cm}  
\caption{ Tensor-isospin terms of the interactions used in our study
as a function of the momentum of the interacting pair.\\~\\~\\~\\~
\label{fig:tens}
}
\end{minipage}
\end{figure}

\vspace*{0.5cm}
In Fig. \ref{fig:tens} we show the momentum dependent part of the 
tensor forces, defined as 
\beq
v_6(q)\,S_{12}(\bq) = 
\int d^3 r \, e^{i\bq\cdot\nr} \,v^{t\tau}(r)\, S_{12}(\nr) 
= - 4 \pi \int dr\,r^2\,j_2(qr)\,v^{t\tau}(r) S_{12}(\nr)  
\,\,\,,
\label{eq:four}
\eeq
where we have indicated with $S_{12}$ the usual expression of the
tensor operator \cite{ari07}.

In this figure, the full line shows the the tensor term of the Argonne
V8' interaction used to build the D1SV8 and D1MV8 forces,
indicated by the dashed-dotted and dashed-doubly dotted lines.  With
the dotted line we show the tensor term of the GT2 interaction
\cite{ots06}. The dashed line has been obtained by multiplying the
bare Argonne V8' term with the scalar part of the short range
correlation calculated in Correlated Basis Function theory
\cite{ari07}. This line indicates that the effect of the short range
correlations obtained by a microscopic calculations are much smaller
than those required by our procedure.  

With the interactions presented above, we have studied a set of
isotopes of doubly magic nuclei. These nuclei have been chosen in such
a way that the single particle levels below the Fermi energy are fully
occupied. In this way we avoided deformation problems and we minimized
pairing effects. We have studied ground states and excitation spectra
of nuclei in the oxygen, calcium, zirconium and tin regions and also
$^{208}$Pb, for a total number of 16 different isotopes.

We have seen that the effects of the tensor force 
are not important on the binding
energies, as already pointed out in Ref. \cite{co98b}. We have
obtained variations of less than 1\% of the values obtained without
tensor terms.  

We have also studied the protons and neutrons density distributions for
all the nuclei mentioned above. Also in this case the effects of the
tensor force are irrelevant. The largest differences between calculations
with and without tensor force are of the order of few parts on a thousand.

\begin{figure}[h]
\hspace*{-0.45cm}
\includegraphics[width=5.8cm,angle=90]{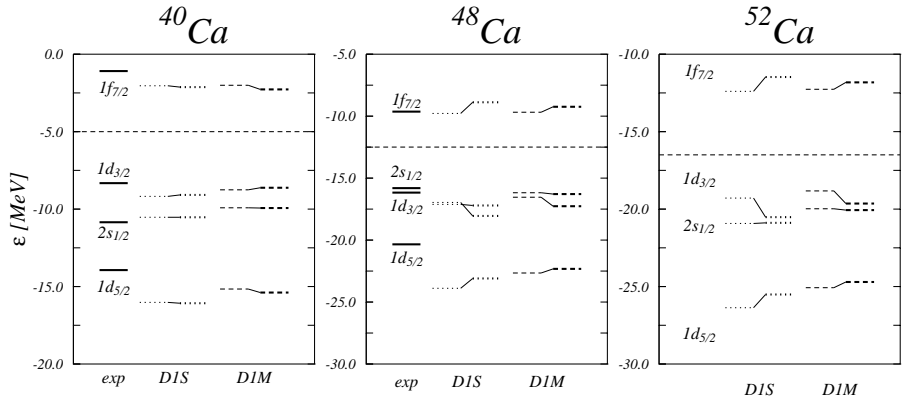}
\hspace{0.2 cm}   
\begin{minipage}[b]{5.0 cm}  
\caption{Single particle energies of the calcium isotopes calculated with the
interactions D1S and D1M, thin lines, and with their corresponding
interactions containing the tensor term, the D1SV8 and D1MV8, thick
lines. \\~\\~\\~\\~
\label{fig:speca}
}
\end{minipage}
\end{figure}

On the opposite, we have observed that the effect of the tensor force on the
single particle energies in not negligible. As example we
show in Fig. \ref{fig:speca} the single particle spectrum of the three
calcium isotopes considered in our study.  For simplicity we present
only the results obtained with D1S and D1M interactions, and their
corresponding interactions containing the tensor term.  In the figure,
the thin lines show the results obtained without tensor force and the thick
lines those obtained with the tensor force. The full lines show the
experimental single particle energies taken from Ref. \cite{dol76}.

As expected, in \caI the tensor term has no effect, since all the
neutron spin-orbit partner levels are occupied.  The small differences
between the energies are due to the different spin-orbit terms between
the D1S and D1SV8 forces and the D1M and D1MV8 forces.

The effect is present in the single particle spectrum of the other two
isotopes since in these nuclei the neutron $j=l+1/2$ levels are
occupied, while their spin-orbit partners are empty.  The figure shows
that with the inclusion of the tensor term the energy of the
1d$_{5/2}$ levels increases, and that of the 1d$_{3/2}$ decreases. For
the \caII nucleus, this effect helps in reproducing the empirical
sequence of single particle states.  For the D1S interaction the
tensor force lowers the energy of the 1d$_{3/2}$ level below that of
the 2s$_{1/2}$ as it is experimentally observed. The effect in
$^{52}$Ca is not so dramatic.  From the experimental point of view it
is not yet clear if the spin and parity of the $^{51}$K ground state
is 3/2$^+$ or 1/2$^+$.  All our calculations indicate 3/2$^+$.

The action of the tensor force affects also the energy of the first proton
level above the Fermi surface, the 1f$_{7/2}$ level, by increasing its
value. This means that the gap between the occupied 1d$_{3/2}$ proton
level and the 1f$_{7/2}$ empty level increases if the tensor force is
included.  For this reason the energy of excited states dominated by
the proton [1f$_{7/2}$,1d$_{3/2}^{-1}$] particle-hole pair should
increase. This is what we observe from the results shown in
Tab. \ref{tab:4-} where we present the RPA eigenergies for the 4$^-$
states in the three calcium isotopes obtained from calculations done
with forces with (D1SV8 and D1MV8) and without (D1S and D1M) tensor force.
We also show results of calculations where the tensor force has been
switched off only in the RPA calculation. They are indicated by the
D1SV8* and D1MV8* labels.

The results behave exactly as expected.  The tensor force has practically no
effect on the \caI results, while there is a clear increase in \caII
and $^{52}$Ca.  On the other hand, the experimental values do not seem
to have this trend. It would be interesting to know the experimental
value of the 4$^-$ energy in $^{52}$Ca. 

%
%
\begin{table}[h]
\centering
\caption{\label{tab:4-}
Energies, in MeV, of the 4$^-$ excited state for various 
Ca isotopes.
}  
\begin{tabular}{lccccccc}
\hline\hline
    &  exp   &  D1S  & D1SV8 & D1SV8* & D1M & D1MV8  & D1MV8* \\
\hline
  $^{40}$Ca &  7.66  &   7.50 & 7.41 & 7.32 & 7.04 & 6.70 & 6.64 \\
  $^{48}$Ca &  6.11  &   7.91 & 8.61 & 8.75 & 6.81 & 7.30 & 7.40 \\ 
  $^{52}$Ca &        &   7.04 & 9.03 & 9.15 & 6.77 & 7.84 & 7.97 \\ 
\hline\hline
\end{tabular}
\end{table}
%
%
\newpage

\begin{figure}[h]
\includegraphics[width=12.8 cm, angle=90,scale=0.7] {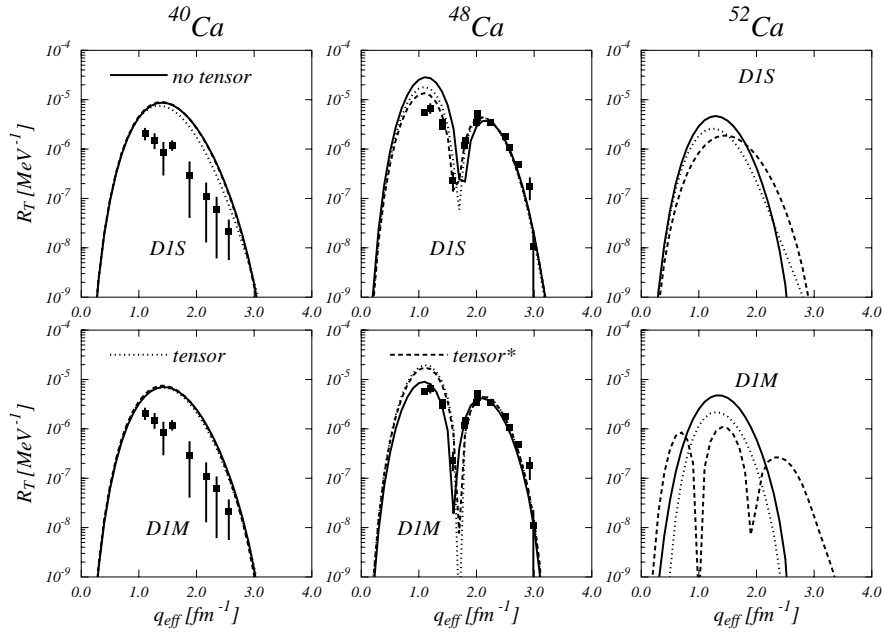} 
\vspace*{-0.6cm} 
\caption{
Electron scattering transverse responses calculated by using
interactions without tensor, full lines, with tensor, dotted lines and
by switching off the tensor force only in the RPA calculations, dashed
lines. The experimental data are taken form Refs. \cite{wil87} and
\cite{wis85}. 
\label{fig:4-}
}
\end{figure}

To have information about the wave function of the excited states, we
have calculated the inelastic electron scattering transverse
responses. In Fig. \ref{fig:4-}, we compare our results with the
experimental data from Ref. \cite{wil87} and \cite{wis85}.

The effects of the tensor force are irrelevant in the \caI case. We observe
some sensitivity to the tensor force in the case of \caII. In the D1S case
the tensor force lowers the first peak, while the effect is reversed in the
case of the D1M interaction. The main effect is that on the single
particle wave functions, on the HF calculation. The situation is
different for the $^{52}$Ca, isotope. In this case the tensor force effects
are relevant on both the HF results, and also on the results obtained
by the RPA calculation. It is interesting the situation of the D1M
calculation in $^{52}$Ca, where the presence of the tensor force in the RPA
eliminates the contamination of the neutron
[2d$_{5/2}$,2p$_{1/2}^{-1}$] particle-hole pair, and it changes the
shape of the response.  

In general, the tensor force does not produce relevant effects on the
bulk properties of the nuclear systems, such as binding energies,
density distributions and excitation energies.  Concerning the excited
states, however, there are situations where the inclusion of the
tensor form is important. Certainly the energies of the 0$^-$ states
are extremely sensitive to the presence of the tensor term of the
force. Furthermore, the effects on the single particle energies can
modify excitations dominated by specific particle-hole pairs. We have
presented the case of the 4$^-$ excitations in calcium isotopes which
clearly shows tensor force effects when the neutrons do not occupy all
the spin-orbit partners levels.

The accuracy required by modern nuclear structure calculations
requires the inclusion of the tensor term in the effective
nucleon-nucleon interaction, especially for the study of nuclei with
neutron excess.



\begin{thebibliography}{10}
\expandafter\ifx\csname url\endcsname\relax
  \def\url#1{{\tt #1}}\fi
\expandafter\ifx\csname urlprefix\endcsname\relax\def\urlprefix{URL }\fi
\providecommand{\eprint}[2][]{\url{#2}}

\bibitem{pud97}
Pudliner B~S, Pandharipande V~R, Carlson J, Pieper S~C and Wiringa R~B 1997
  {\em Phys.\ Rev. \ C\/} {\bf 56} 1720

\bibitem{don09}
De~Donno V, Co' G, Maieron C, Anguiano M, Lallena A~M and Moreno-Torres M 2009
  {\em Phys. \ Rev. \ C\/} {\bf 79} 044311

\bibitem{led78}
Lederer C~M and Shirley V~S 1978 {\em Table of isotopes, 7th ed.\/} (John Wiley
  and sons, New York)

\bibitem{ots05}
Otsuka T, Suzuki T, Fujimoro R, Grawe H and Akaishi Y 2005 {\em Phys. \ Rev. \
  Lett.\/} {\bf 95} 232502

\bibitem{ber91}
Berger J~F, Girod M and Gogny D 1991 {\em Comp. \ Phys. \ Comm.\/} {\bf 63} 365

\bibitem{gor09}
Goriely S, Girod M, Hilaire S and P\'eru S 2009 {\em Phys. \ Rev. \ Lett.\/}
  {\bf 102} 252501

\bibitem{ots06}
Otsuka T, Matsuo T and Abe D 2006 {\em Phys. \ Rev. \ Lett.\/} {\bf 97} 162501

\bibitem{ari07}
Arias~de~Saavedra F, Bisconti C, Co' G and Fabrocini A 2007 {\em Phys. \ Rep.\/}
  {\bf 450} 1

\bibitem{co98b}
Co' G and Lallena A~M 1998 {\em Nuov. \ Cim. \ A\/} {\bf 111} 527

\bibitem{dol76}
Doll P, Wagner G~J, Kn{\"o}pfle K~T and Mairle G 1976 {\em Nucl. \ Phys. \ A\/}
  {\bf 263} 210

\bibitem{wil87}
Williamson C~F and et~al 1987 unpublished

\bibitem{wis85}
Wise J and et~al 1985 {\em Phys. \ Rev. \ C\/} {\bf 31} 1699

\end{thebibliography}
\end{document}